\documentclass[runningheads]{svmult}

\usepackage{makeidx}   % allows index generation
\usepackage{graphicx}  % standard LaTeX graphics tool
\usepackage{subeqnar}  % subnumbers individual equations
\usepackage{multicol}  % used for the two-column index
\usepackage{physprbb}  % modified textarea for proceedings,
\makeindex             % used for the subject index
\begin{document}
\title*{The MUNICS Project: Galaxy Assembly at $0 < z < 1$}
\toctitle{The MUNICS Project: Galaxy Assembly at $0 < z < 1$}
% allows explicit linebreak for the table of content
%
%
\titlerunning{The MUNICS Project}
% allows abbreviation of title, if the full title is too long
% to fit in the running head
%
\author{Niv Drory\inst{1}
\and Ralf Bender\inst{2,3}
\and Georg Feulner\inst{3}
\and Gary J.\ Hill\inst{1}
\and Ulrich Hopp\inst{3}
\and Claudia Maraston\inst{2}
\and Jan Snigula\inst{3}}
\authorrunning{Niv Drory et al.}
% if there are more than two authors,
% please abbreviate author list for running head
%
%
\institute{University of Texas at Austin, Austin, Texas 78712
\and Max--Planck Institut f\"ur extraterrestrische Physik,
  Giessenbachstra\ss e, Garching, Germany
\and Universit\"ats--Sternwarte M\"unchen, Scheinerstra\ss e 1, 
D-81679 M\"unchen, Germany}

\maketitle              % typesets the title of the contribution

\section{The MUNICS Survey and its Results}
MUNICS is a wide-area, medium-deep, photometric and spectroscopic
survey selected in the K band, targeting randomly-selected high
Galactic latitude fields. It covers an area of roughly one square
degree in the K and J bands with complementary optical follow-up
imaging in the I, R, V, and B bands in 0.5 square degrees.

The limiting magnitudes of this main part of the survey are 19.5 in K,
21.5 in J, 22.5 in I, and 23.5 in R (50 \% completeness for point-like
sources). This multicolor catalog probes field galaxies in a large
volume out to redshifts of roughly 1.5 (for massive galaxies) and is
by far the largest catalog of near-infrared selected distant galaxies
published so far. It thus comprises a suitable and highly competitive
multi-color field galaxy survey.  The photometric survey is described
and characterized in \cite{MUNICS1} and \cite{MUNICS4}.  The survey
spans the redshift range $0<z<1.5$ and selects typically $L^*$ and
brighter objects.

The MUNICS photometric survey is complemented by spectroscopic
follow-up observations at 4m-class telescopes of all galaxies down to
$K\le 17.5$ in 0.25 square degrees. This survey is complete down to
$K\le 16.5$ and 80\% complete at $16.5 < K < 17.5$ to the present
date. Furthermore, a sparsely selected deeper sample down to $K \le
19$ was observed with the ESO VLT, covering a much smaller area of 100
square arcmin. The whole spectroscopic sample contains 593 secured
redshifts thsu far. The spectra cover a wide wavelength range
$4000-8500$\AA\ at $13.2$\AA\ resolution, and sample galaxies at
$0<z<1$. Details of the spectroscopic survey are given in
\cite{MUNICS5}.

Since it is impossible to obtain spectroscopic redshifts for the whole
MUNICS sample (most galaxies in the sample are too faint to be
observed currently with optical spectrographs) one must rely on
photometric redshift techniques to obtain distance estimates for the
complete sample of almost 6000 galaxies. Comparing to $> 500$
spectroscopic redshifts, the scatter in the relative redshift error
$\Delta z / (1+z)$ is 0.055. The distribution of photometric redshifts
peaks around $z \approx 0.5$. and has a tail extending to $z \approx
3$.

Given these limits, MUNICS contains mostly massive (stellar mass $M >
10^{10} M_\odot$) field galaxies, spanning a significant fraction of
cosmic time and is ideal for studying their formation and evolution,
specifically their mass assembly history.

\section{The Rest-frame K-Band Luminosity Function to $\bf z \sim 1.2$}

\begin{figure}
  \centering
  \includegraphics[width=0.6\textwidth]{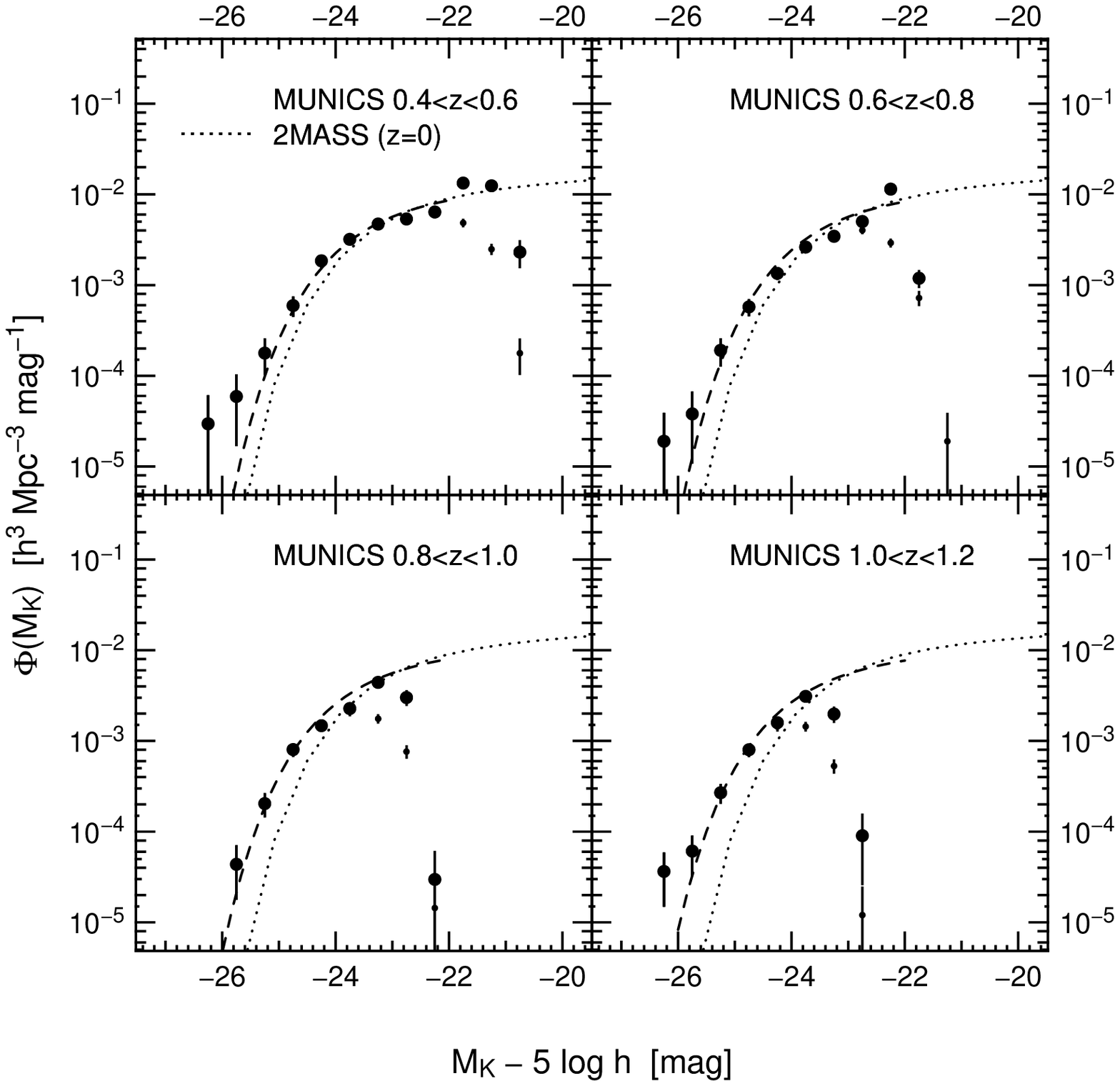}
  \includegraphics[width=0.38\textwidth]{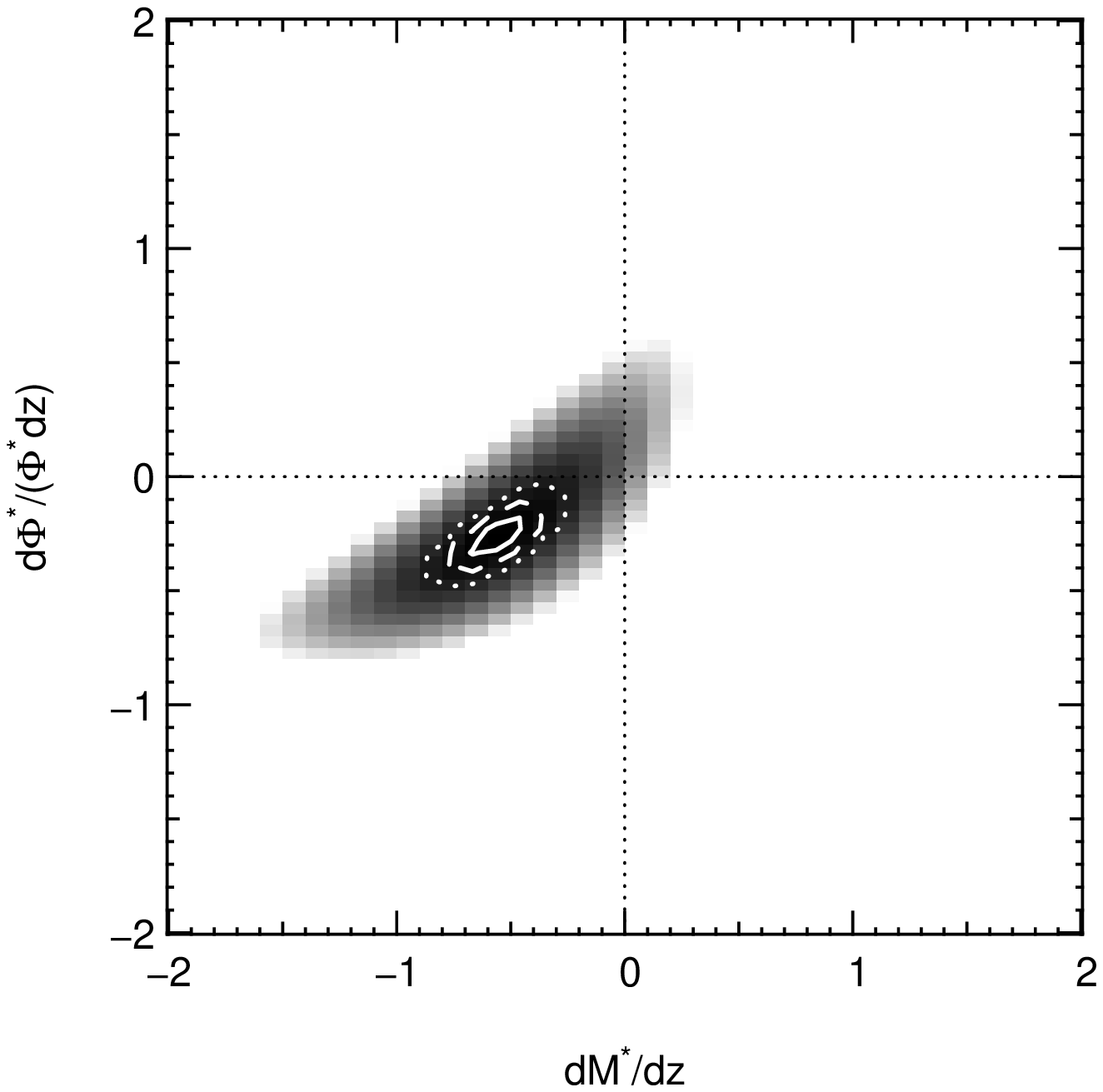}\hfill
  \caption{\label{f:lf}
    {\bf Left panel:} The rest-frame K--band luminosity function
    derived from the MUNICS data in four redshift bins spanning $0.4 <
    z <1.2$.  The dotted curve is the $z=0$ LF 
    \cite{Kochaneketal01}. The dashed lines are parameterized fits to
    the data. {\bf Right panel:} Likelihood map for the change in the
    Schechter parameters, $d\Phi^*/(\Phi^*\,dz)$ and $dM^*/dz$, with
    redshift.}
\end{figure}

Fig.\ \ref{f:lf} shows the rest-frame K--band luminosity function
(LF) derived from the MUNICS data in four redshift bins spanning $0.4
< z <1.2$ \cite{MUNICS2}. Absolute magnitudes were derived using the
photometric redshifts, extrapolating the best fitting SED to the
rest-frame K band. Errors due to this extrapolation are expected to
be small in the redshift range we probe, since the near-IR slopes of
the SEDs differ only very little over the galaxy types probed by our
survey (the K--band $k$-corrections are small and almost
type-independent).

\begin{figure}
  \centering
  \includegraphics[width=0.5\textwidth]{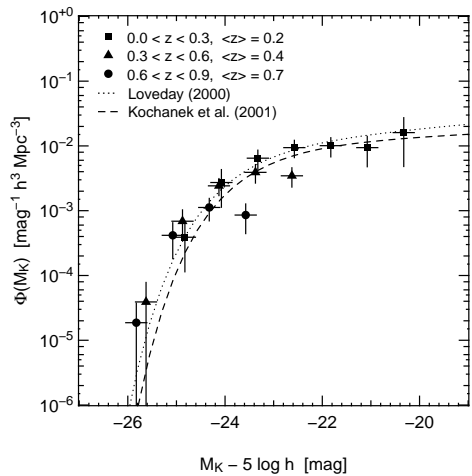}\hfill
  \caption{\label{f:slf}
    The K band luminosity function of galaxies from spectroscopic
    observations of MUNICS galaxies. The dotted and dashed curves are
    the $z=0$ LFs \cite{Loveday00,Kochaneketal01}.}
\end{figure}

Quantitative analysis of the evolution of the LF yields a mild
decrease in number density by $\sim 25\%$ to $z=1$ accompanied by
brightening of the galaxy population by $\sim 0.5$ mag. These results
are fully consistent with an analogous analysis using only the
spectroscopic MUNICS sample \cite{MUNICS5}, which is shown in
Fig.~\ref{f:slf}.

To interpret these results in terms of a picture of galaxy evolution
is not straight--forward, though. Since the K--band light traces
stellar mass, we may attribute a change in K band luminosity function
either to a change in the mass--to--light ratio alone (passive
evolution) or to a combination of a change in $M/L$ and in stellar
mass. A change in stellar mass can be due to star formation and/or due
to merging and accretion. These models cannot be discriminated on
basis of the luminosity function alone. In fact, if the increase seen
in the global star formation rate to $z \sim 1$ (e.g.\ 
\cite{CFRS96,CFRS1297,HCBP98,CSB99}) is attributable to normal field
galaxies, the stellar mass of these systems is expected to evolve by
roughly a factor of 1.5 to 2 over the redshift range $0<z<1$.
Therefore, if the K band light does reflect stellar mass, number
density evolution in the K band is indeed expected.

\section{The Stellar Mass of Field Galaxies to $\bf z \sim 1.2$}

The multicolor data in MUNICS with their wide range in wavelength
allow us to investigate properties of the stellar populations of
individual objects in greater detail. We can therefore use the MUNICS
data to derive the stellar mass function in the redshift range $0.4 <
z < 1.2$ \cite{MUNICS6,MUNICS3} by modeling and fitting the stellar
mass-to-light ratios in the NIR. 

We parameterize star formation histories as $\psi(t) \propto
\exp(-t/\tau)$, with $\tau \in \{0.1, 0.2, 0.4, 1.0, 2.0, 3.0, 5.0,
8.0, 10.0, 13.0\}$~Gyr. We extract spectra at 28 ages between 0.001
and 14~Gyr and allow $A_v$ to vary between 0 and 3 mag using a
Calzetti \cite{Calzetti00} extinction law. The models use solar
metallicity and are based on the Simple Stellar Population models by
C.~Maraston \cite{Maraston98}. We assume a value of 3.33 for the
absolute K--band magnitude of the Sun. We use a Salpeter IMF with
lower and upper mass cutoffs of 0.1 and 100~$M_\odot$. This choice
allows us to compare our results directly with the literature. The use
of an IMF with a flatter slope at the low mass end will not affect the
shape of the mass function, it will only change its overall
normalization. If, however, the IMF depends on the mode of star
formation, e.g.\ being top--heavy in starbursts, our results will be
affected. We convert the absolute K--band magnitude, $M_K$, into
stellar mass by using the K--band mass--to--light ratio, $M/L_K$, of
the best fitting CSP model in a $\chi^2$\ sense.

\begin{figure}
  \centering \includegraphics[width=0.7\textwidth]{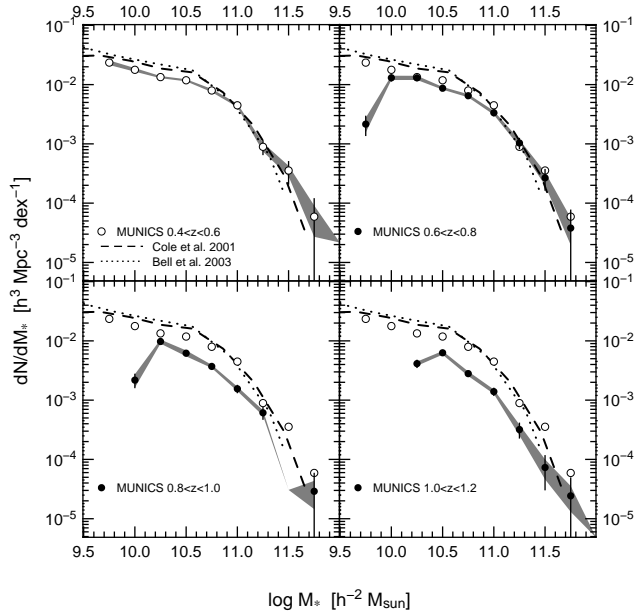}
  \caption{\label{f:mf}%
    The evolution of the stellar mass function with redshift. The open
    symbols are the MUNICS values at $0.4 < z < 0.6$, the closed
    symbols are the MUNICS values at higher redshifts. The lowest $z$
    values are shown in all panels for comparison. Error bars denote
    the uncertainty due to Poisson statistics. The shaded areas show
    the 1~$\sigma$ range of variation in the mass function given the
    total systematic uncertainty in $M/L_K$. The dotted and dashed
    lines show the $z=0$ stellar mass function derived similarly to
    our methods using SDSS, 2dF, and 2MASS data \cite{BMKW03,2dF01}.}
\end{figure}

Fig.~\ref{f:mf} shows the mass function of galaxies with $10^{10} <
M/(h^{-2}\,M_\odot) < 10^{12}$ in four redshift bins centered at $z=0.5$,
$z=0.7$, $z=0.9$, and $z=1.1$. The results on the local stellar mass
function \cite{2dF01,BMKW03} derived by fitting stellar population
models to multicolor photometry and deriving NIR mass--to--light
ratios similarly to the method employed here are also shown for
comparison as a dashed and dotted line, respectively. The data from
the lowest redshift bin, $0.4 < z < 0.6$ are shown alongside the
higher--redshift data for easier comparison as open symbols. Error
bars denote the uncertainty due to Poisson statistics. The shaded
areas show the 1~$\sigma$ range of variation in the mass function from
Monte--Carlo simulations given our estimate of the total systematic
uncertainty in $M/L_K$.

The lowest redshift bin shows remarkable agreement with the $z=0$
values, despite the different selection at low versus high redshift
and the different model grids used, although we obtain slightly lower
number densities at $\log M/(h^{-2}\,M_\odot) \leq 10.5$. Therefore, there
seems to be not much evolution in stellar mass at $z < 0.5$. The
general trend at higher redshift, is for the total normalization of
the stellar mass function to goes down and for the knee to move
towards lower masses. This causes the higher masses to evolve faster
in number density than lower masses and is well visible in
Fig.~\ref{f:mf} at $10.5 < \log M/(h^{-2}\,M_\odot) < 11.5$.

\section{The total stellar mass density}

\begin{figure}
  \centering \includegraphics[width=0.6\textwidth]{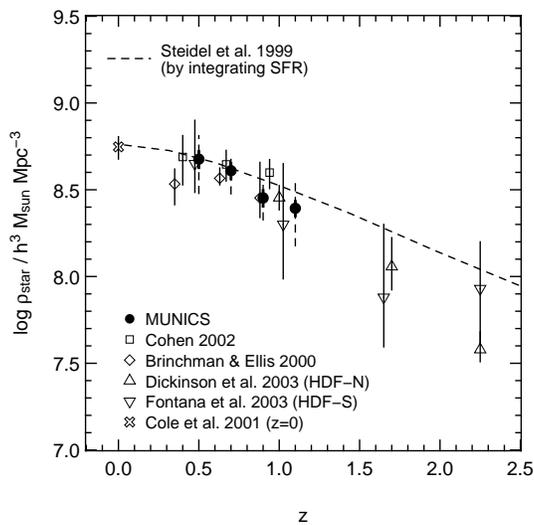}
  \caption{\label{f:md}%
    The evolution of the total stellar mass density in the universe.
    The closed circles are the MUNICS values, open symbols are values
    from the literature. The integrated star formation rate (dashed
    curve) is shown for comparison. The thick error bars on the MUNICS
    values (solid dots) are the statistical errors associated with the
    data. The dashed error bars show the variance we get from the
    MUNICS data in GOODS size patches.}
\end{figure}

The total stellar mass density of the universe -- the integral of the
star formation history of the universe (e.g.\ 
\cite{Madauetal96,SAGDP99}) and observationally its complement -- is
shown in Fig.~\ref{f:md}. We also plot the local value from
\cite{2dF01}, values from the Hubble Deep Fields
\cite{DPFB03,Fontanaetal03} covering $z > 1$ and values from
\cite{Cohen02,BE00} at $z < 1$. Additionally, we integrate the star
formation history curve (including extinction correction) from
\cite{SAGDP99} for comparison.

\clearpage

From our data, as well as from the HDFs and the integrated star
formation rate from the UV luminosity density, it appears that 50\% of
the local mass in stars has formed since $z \sim 1$. The data from
\cite{BE00} are consistent with ours at $z \sim 1$ but are lower at $z
\sim 0.4$ and seem to under--predict the local value if extrapolated.
The values obtained by \cite{Cohen02} are higher at $z \sim 1$ which
we think is attributable to their method of deriving $M/L_K$.

It is worth noting that the results from the HDF--N differ by a factor
$\sim 2$ from the HDF--S, attributable to cosmic variance.  The MUNICS
values in Fig.~\ref{f:md} are shown with their statistical errors
(thick error bars), which amount to roughly 10\%. We also show the
variance we get in our sample divided into GOODS size patches of 150
square arcminutes (dashed error bars), showing that at these
redshifts, even surveys like GOODS are expected to be dominated by
cosmic variance. We expect differences of around 50\% between the two
GOODS areas.

%INDEX%%%%%%%%%%%%%%%%%%%%%%%%%%%%%%%%%%%%%%%%%%%%%%%%%%%%%%%%%%%%%%%
% Please check with the editor of your book whether he plans to
% include a "mutual" subject index - if so, please code your entries
% in the standard syntax. For your own purposes you may print your
% "personal" index by using the following commands:
%
%\clearpage
%\addcontentsline{toc}{section}{Index}
%\flushbottom
%\printindex
%%%%%%%%%%%%%%%%%%%%%%%%%%%%%%%%%%%%%%%%%%%%%%%%%%%%%%%%%%%%%%%%%%%%%

\end{document}